\documentclass[11pt]{article}
\usepackage{graphicx}
\newcommand{\be}{\begin{equation}}
\newcommand{\ee}{\end{equation}}
\newcommand{\la}{\langle}
\newcommand{\ra}{\rangle}
\newcommand{\mD}{\mathcal{D}}
\newcommand{\Tr}{\mbox{Tr}}

\begin{document}

\Large
\begin{center}
Scalar Particles in Lattice QCD\footnote{Talk presented at
``International Symposium on Hadron Spectroscopy, Chiral Symmetry
and Relativistic Description of Bound Systems'', Feb. 2003, Tokyo.}
\end{center}
\normalsize

\begin{center}
SCALAR Collaboration  
\\
Teiji Kunihiro$^{a)}$,
Shin Muroya$^{b)}$, Atushi Nakamura$^{c)}$, Chiho Nonaka$^{d)}$,
Motoo Sekiguchi$^{e)}$ and Hiroaki Wada$^{f)}$ \\
\end{center}

\begin{center}
\begin{small}
{\it 
$^{a)}$ YITP, Kyoto University, Kyoto 606-8502, Japan,
\\
$^{b)}$ Tokuyama Women's College, Tokuyama 745-8511, Japan, 
\\
$^{c})$ RIISE, Hiroshima University, Higashi-Hiroshima 739-8521, Japan, 
\\
$^{d})$Department of Physics, Duke University, Durham, NC 27708-0305, USA,
\\
$^{e)}$ Faculty of Engineering, Kokushikan University, Tokyo 154-8515,
Japan,  
\\
$^{f)}$ Laboratory of Physics, College of Science and Technology,\\
Nihon University, Chiba 274-8501, Japan.}
\end{small}
\end{center}

\begin{abstract}
We report a project to study scalar particles by lattice
QCD simulations.
After a brief introduction of the current situation of 
lattice study of the sigma meson, we describe our numerical
simulations of scalar mesons, $\sigma$ and $\kappa$.
We observe a low sigma mass, $m_\pi<m_\sigma\le m_\rho$, 
for which the disconnected diagram plays an important role. 
For the kappa meson, we obtain higher mass than the experimental 
value, {\it i.e.} $m_\kappa\sim 2m_{K^*}$. 
\end{abstract}


\section{Introduction}

The objective of our collaboration is to understand scalar mesons
in the framework of QCD.
The confidence level of the sigma meson has been increasing, and
an other scalar meson \cite{KEK,close-tornqvist}, 
$\kappa$, has been reported by several experimental groups.
In the modern hadron physics based on QCD, the chiral symmetry is an
important ingredient and the sigma meson plays an essential role in it
together with the pion. 

The existence of the sigma meson was obscure for many years. 
The re-analyses of the $\pi$-$\pi$ scattering phase shift have 
suggested a pole of the $\sigma$  meson with $I=0$ 
and $J^{PC}=0^{++}$\cite{crossing}.
In this analysis, the chiral symmetry, analyticity, unitarity 
and crossing symmetry are taken into account. 
Contributions of the $\sigma$ pole were observed in the decay processes
such as D$\to \pi \pi \pi $\cite{decay} 
and $\Upsilon(3S) \to \Upsilon \pi \pi$ \cite{Ishida}.
In  1996 PDG(Particle Data Group), ``$f_0$(400-1200) or $\sigma$" 
appeared below $1$ GeV mass region, and ``$f_0$(600) or $\sigma$"  
in the 2002.

If the sigma meson exists, it is natural to consider
the $\kappa$ meson as a member of the nonet scalar states of chiral 
SU(3)$\otimes$SU(3) symmetry. Recently,
the $\kappa$ with $I=1/2$ is reported with mass $m_{\kappa}\sim$ 
800 MeV \cite{E791,BES}. 

We believe that it is very important now to investigate these scalar mesons
by lattice QCD in order to establish scalar meson
spectroscopy as a sound and important piece of hadron physics.
Lattice QCD provides a first-principle approach of hadron physics,
and allows us to study non-perturbative aspects of quark-gluon dynamics.
It is a relativistic formulation, and 
quarks are described as Dirac fermions.
It is {\it not} a model, and apart from numerical limitations, 
there are no approximations.
It is not a bound state calculation: neither a potential model nor 
Bethe-Salpeter calculation. 

Lattice QCD is usually formulated in the Euclidean path integral as
\be
Z  = \int \mD U \mD\bar{\psi}\mD\psi
    \, e^{- S_G - \bar{\psi} D \psi}
 = \int \mD U \det D \, e^{- S_G} ,
\label{Eq:PathInte}
\ee 
where $S_G$ is the gluon kinetic action, whose continuum limit is
$
-\int d^4 x \Tr F_{\mu\nu}^2 / 4
$
We construct a state with a definite quantum number and measure the decay
in the channel,
\be
G(x,y) = \frac{1}{Z}\int\mD U \mD\bar{\psi}\mD\psi H(y)H(x)^\dagger 
         \, e^{- S_G - \bar{\psi} D \psi}
       = \la H(y)H(x)^\dagger \ra
         \rightarrow
         e^{-m|x-y|}
\ee
where $H(x)$ is a hadron operator.  For scalar mesons, 
$
H(x) = 
         \bar{\psi}^{a f_1}_\nu \Gamma_{f_1 f_2}
         \psi^{a f_2}_\nu .
$
The indices $a$, $f_i$ and $\nu$ stand for color, flavor and Dirac indices, 
respectively.
$H(x)^\dagger |0\ra $ is a state whose quantum number is specified by the 
operator $H$.

We must be very careful of the limitation of the present
lattice QCD calculations:
\begin{enumerate}
\item
Sufficient statistics:
Gauge configurations are generated by the Monte Carlo simulation, 
and there are statistical errors like experimental data.
\item
Continuum limit: Numerical simulations are performed at a finite
lattice spacing, $a$, and we must take $a \to 0$ limit at the end. 
\item
Infinite volume limit:
Lattice volume should be large enough to include a hadron. 
\item
Chiral extrapolation:
$u$ and $d$ quark masses on the current lattice are still large
and extrapolated to zero.
\end{enumerate}
The last point may cause systematic bias.  $\pi$ mesons are not 
sufficiently light and our sigma meson cannot decay into 2 $\pi$, 
{\it i.e.}, its width is zero.
Any other particle, {\it e.g.} $\rho$, $\Delta$ and $N^*$, also 
has zero width in lattice QCD calculations in the literature.
In the case of the sigma meson, this flaw should be kept in mind, 
since the two-pion component may be important.

There have been several attempts at lattice study of sigma mesons.
To our knowledge, the first such calculation was carried out by 
deTar and Kogut \cite{DeTar}, 
where the so-called disconnected diagram, or the OZI forbidden type 
diagram is discarded.  
The channel was called `valence' sigma, $\sigma_V$.
They measured the screening masses and observed that $\sigma_V$ is 
much heavier than the $\pi$ meson at the zero temperature, 
while $\sigma_V$ and $\pi$ degenerate as the temperature increases over $T_c$.
Kim and Ohta calculated the valence sigma mass with staggered fermions
for the lattice spacing $a=0.054$ fm 
and lattice size $48a=2.6$ fm \cite{KimOhta}.
They obtained $m_\sigma/m_\pi = 1.4 \sim 1.6$ by varying $m_\pi/m_\rho
=0.65 \sim 0.3$.

Lee and Weingarten \cite{Lee} have stressed the importance of 
the mixing the scalar meson and glueballs and concluded that 
$f_0$(1710) is the lightest scalar glueball dominant particle, 
while $f_0$(1390) is composed of mainly the $u$ and $d$ quarkonium.
Alford and Jaffe analyzed the possibility that the sigma particle is
an exotic state, i.e., $qq\bar{q}\bar{q}$ by a lattice QCD 
calculation \cite{Alford}. 

All these calculations are in the quench approximation, {\it i.e.}, 
the fermion determinant in Eq.(\ref{Eq:PathInte}) is dropped, which
corresponds to ignoring quark pair creation and annihilation diagrams.
McNeile and Michael observed that the $\sigma$ meson masses in the
quench approximation and in the full QCD simulation are very 
different \cite{McNeile}.
They also considered the mixing with the glueballs.
They obtained a very small sigma meson mass, even smaller than
the $\pi$ mass, in the full QCD case.

There are two ongoing projects of $\sigma$ meson spectroscopy :
Riken-Brook haven-Columbia (RBC) collaboration \cite{Sasa}
and Scalar collaboration \cite{Lat01,Lat02}.
The two approaches are complementary.  
The RBC collaboration employs
the domain wall fermions, which respect the chiral symmetry, 
but include a quench approximation, 
while Scalar collaboration uses Wilson fermions, which break the
chiral symmetry at a finite lattice spacing, 
but performs the full QCD simulation.
RBC reported their simulation at $a^{-1}=1.3$ GeV 
on a $16^3\times 32$ lattice.
They remedied the quench defect with the help of the chiral perturbation.
They observed that the masses of the non-isosinglet scalar ($a_0$) and
the singlet scalar ($\sigma$) are almost degenerate 
when the quark mass is heavy (above $s$ quark mass regions), 
and that as the quark mass decreases,
the $a_0$ mass remains almost constant, 
but the mass of the $\sigma$ decreases.  

\section{Propagators}

The quantum numbers of the $\sigma$ meson are $I=0$ and 
$J^{PC}=0^{++}$; we adopt the $\sigma$ meson operator of 
\begin{equation}
\hat{\sigma}(x) \equiv 
\frac{\bar{u}(x)u(x)+\bar{d}(x)d(x)}
{\sqrt{2}} ,
\label{eq:sigma_operator}
\end{equation}
where $u$ and $d$ indicate the corresponding quark spinors, and
we suppress the color and Dirac indices. 
The $\sigma$ meson propagator is written as
\begin{eqnarray}
G_\sigma(y,x) 
&=& \frac{1}{Z} \int \mD U\mD \bar{u}\mD u\mD \bar{d}\mD d \,
\left( \hat{\sigma}(y) \hat{\sigma}(x)^\dagger \right)
e^{-S_G-\bar{u}Du-\bar{d}Dd} . 
\label{eq:propa_def}
\end{eqnarray}
By integrating over $u$, $\bar{u}$, $d$ and $\bar{d}$ fields, 
the $\sigma$ meson propagator is given by 
\begin{eqnarray}
G_\sigma(y,x) 
& = -& \la \Tr D^{-1}(x,y) D^{-1}(y,x) \ra 
\nonumber \\
  & &  + 2 \la ( \sigma(y) - \la \sigma(y) \ra )
                ( \sigma(x) - \la \sigma(x) \ra ) \ra
\label{eq:propa}
\end{eqnarray}
where 
$
\sigma(x) \equiv \Tr D^{-1}(x,x) .
$
``Tr'' represents a summation over color and Dirac spinor indices. 
In Eq.(\ref{eq:propa}), $D^{-1}$'s are $u$ and $d$ quark propagators.
Here we assumed that the $u$ and $d$ quark propagators are equivalent because 
$u$ and $d$ quark masses are almost the same. From Eq.(\ref{eq:propa}), 
we can see that the $\sigma$ propagator consists of two terms. 
The first term corresponds to the connected diagram, 
{\it i.e.}, a $\bar{q}q$-type meson. 

The second term is the ``disconnected'' diagram; 
it is the correlation of $\sigma=\Tr D^{-1}$ at two points $x$ and $y$. 
The term ``disconnected'' is not appropriate since the corresponding
matrix element is, of course, not factorized; quark lines are connected
by gluon interactions.  Nevertheless, we use this jargon in the following. 

The quantum number of the $\sigma$ meson ($I=0$ and $J^P=0^+$) 
is the same as that of the vacuum, and the vacuum expectation 
value of the $\sigma$ operator, $\la \sigma(x) \ra$, does not vanish.  
Therefore, the contribution of $\la \sigma(x) \ra$ 
should be subtracted from the $\sigma$ operator.

For $\kappa^+$, the operator is
$
H(x) = \sum_{a} \sum_{\nu} \bar{s}_\nu^a u_\nu^a ,
$
and we have only the connected diagram,
\be
G_\kappa(y,x)  =  
- \la \Tr D_s^{-1}(x,y) D^{-1}(y,x) \ra ,
\label{Eq:propa-kappa}
\ee
where $D_s^{-1}$ is the $s$ quark propagator.

\section{Numerical simulations}

In this project, we intend to use standard well-established techniques
for numerical calculations, and want to see the outcome.
We employ Wilson fermions and the plaquette gauge action. 
Full QCD simulations were done by the Hybrid Monte Carlo (HMC) algorithm.

CP-PACS performed a very large-scale simulation of light meson 
spectroscopy in the full QCD calculation \cite{CPPACS}. 
We use here the same values of the simulation parameters, 
{\it i.e.},  $\beta = 4.8$ and $\kappa = 0.1846$, 
$0.1874$, $0.1891$, except lattice size;  
our lattice, $8^3\times16$, is smaller.
We employ the point source and sink; smaller lattice size
leads to larger mass due to a higher state mixture.  In other words,
our mass values on the small size lattice should be considered as the
upper limit. 
We  have checked that
the values of $m_\pi$ and $m_\rho$ are consistent with those of CP-PACS.

\begin{figure}[thb]
\begin{center}
\includegraphics[width=.8 \linewidth]{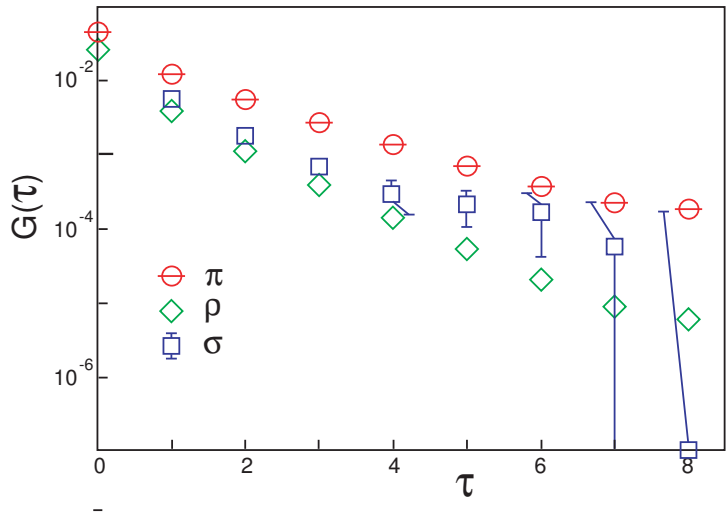} 
\caption{Propagators of $\pi$, $\rho$ and $\sigma$ for 
$\kappa=0.1891$.  }
\label{fig:Propa_k1891}
\end{center}
\end{figure}
\vspace{-0.5cm}

\begin{figure}[thb]
\begin{center}
\includegraphics[width=.7 \linewidth]{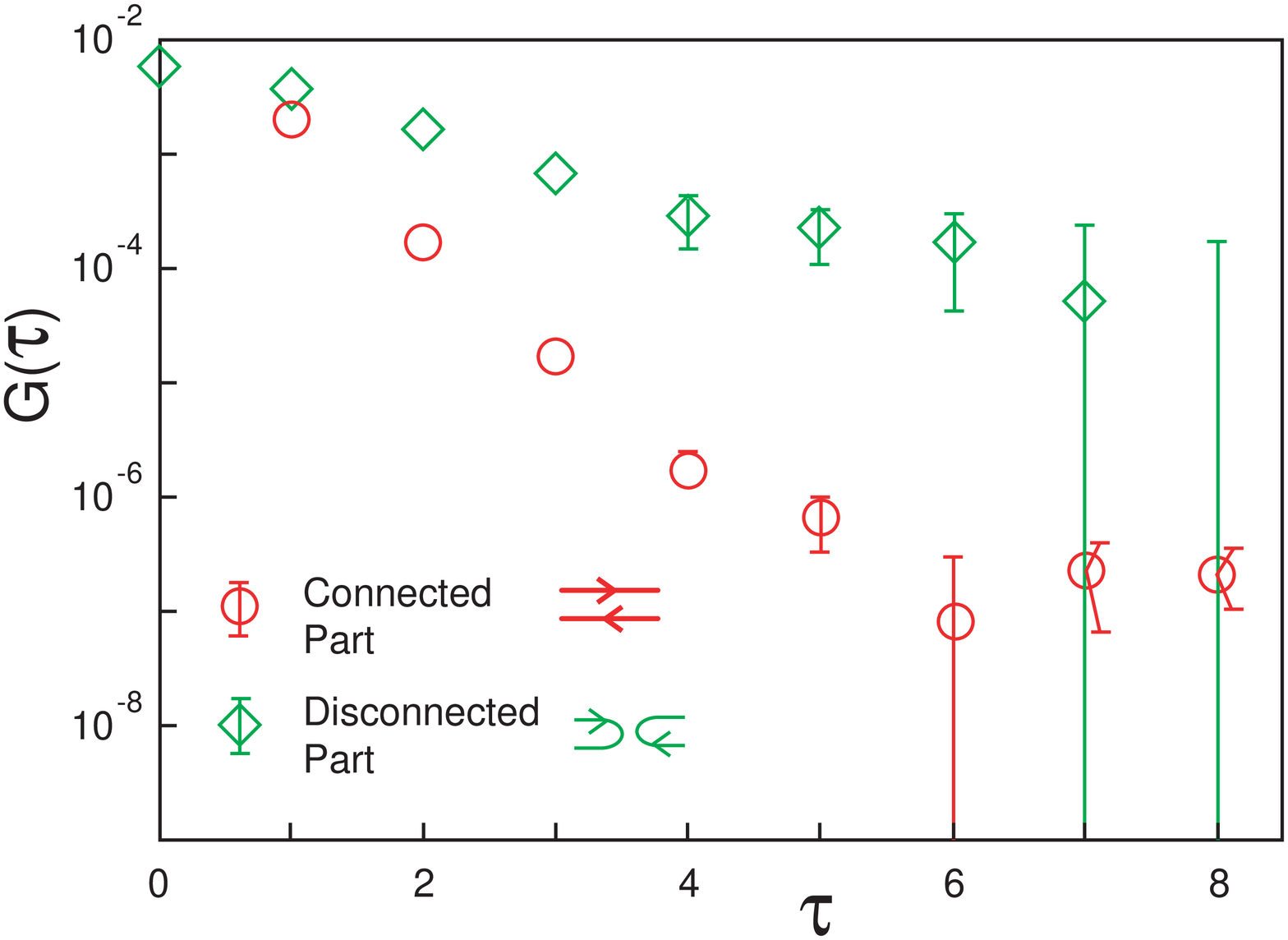} 
\caption{Propagators of the connected and disconnected diagram of 
$\sigma$ for $\kappa=0.1891$.}
\label{fig:Propa_ConnDis}
\end{center}
\end{figure}

It is very difficult to evaluate the disconnected part of the propagator,
since we must calculate $\mbox{Tr}D^{-1}(x,x)$ for all lattice sites $x$.
We used the $Z_2$ noise method to calculate the disconnected diagrams 
and the subtraction terms of the vacuum $\la \sigma \ra$.
Each of these terms is of the order of ten, and 
$\la (\sigma-\la\sigma\ra)(\sigma-\la\sigma\ra)\ra$ becomes less than
$10^{-4}$, as shown in Fig.\ref{fig:Propa_ConnDis}.
Therefore, high accuracy is required for the calculation. 
One thousand random $Z_2$ numbers are generated. 
Our numerical results show that the values of the first  
and the second terms in Eq.(\ref{eq:propa}) are of the same order.  
Therefore, in order to obtain the signal  correctly 
as the difference between these terms, 
high-precision numerical simulations and careful analyses are required. 
We have investigated the relationship between the amount of $Z_2$ noise 
and the achieved accuracy in Ref.\cite{Lat01}.
Gauge configurations were created by HMC in the SX5 vector supercomputer,
and most disconnected propagator calculations by the $Z_2$ noise method
were mainly performed on the SR8000 parallel machine at KEK. 

The propagators of $\pi$, $\rho$ and $\sigma$ for $\kappa=0.1891$ 
are shown in Fig.\ref{fig:Propa_k1891}.
The connected and disconnected parts of the $\sigma$ propagator 
are shown in Fig.\ref{fig:Propa_ConnDis}.
It is difficult to obtain
$\sigma$ propagator at large $\tau$ 
since the precision of our calculation is limited to
{\it O}($G(\tau))\sim10^{-4}$.

\vspace{-0.6cm}

\begin{figure}[thb]
\begin{center}
\begin{minipage}{ 0.47\linewidth}
\includegraphics[width=1.2 \linewidth]{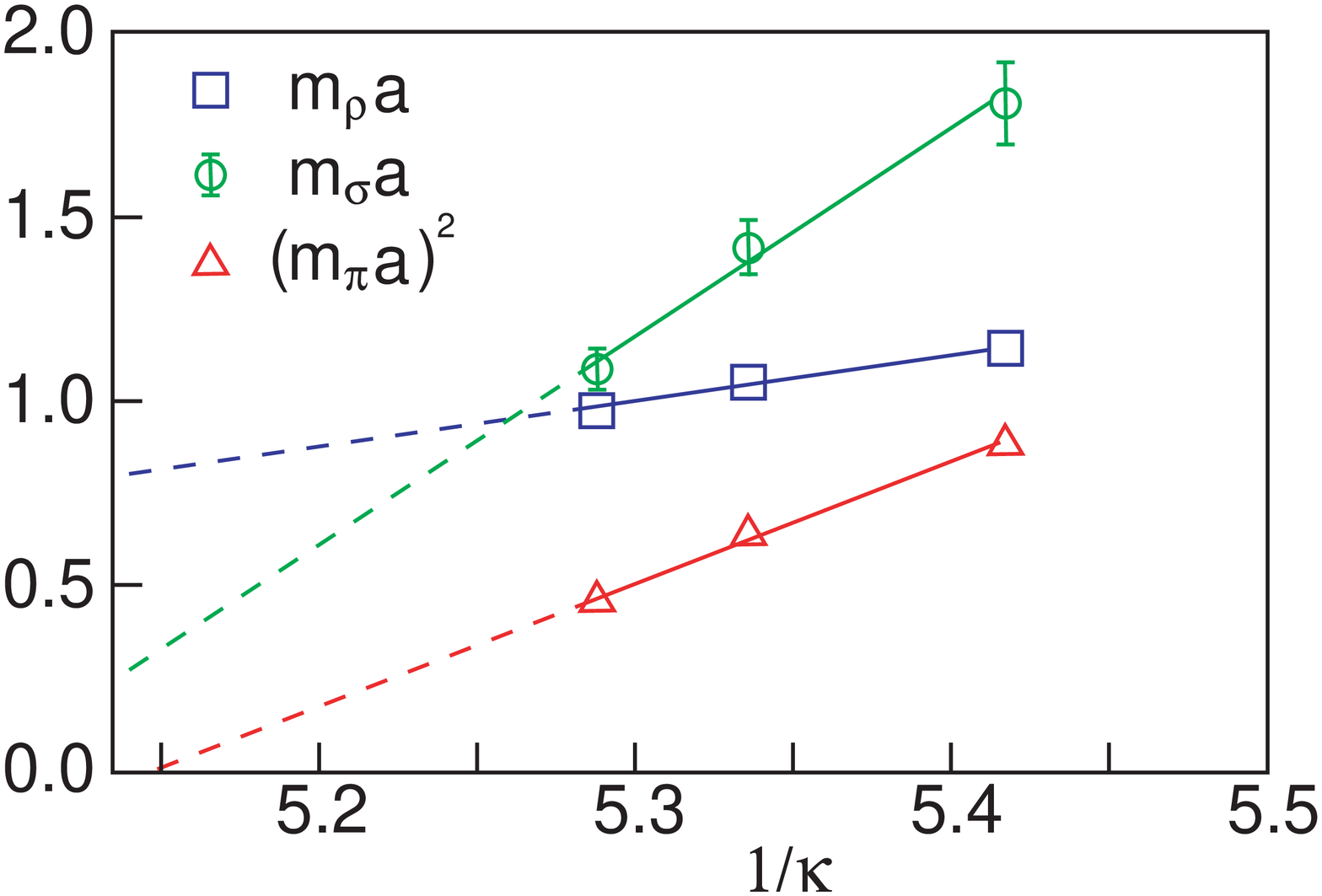} 
\end{minipage}
\hspace{1mm}
\begin{minipage}{ 0.47\linewidth}
\includegraphics[width=1.1 \linewidth]{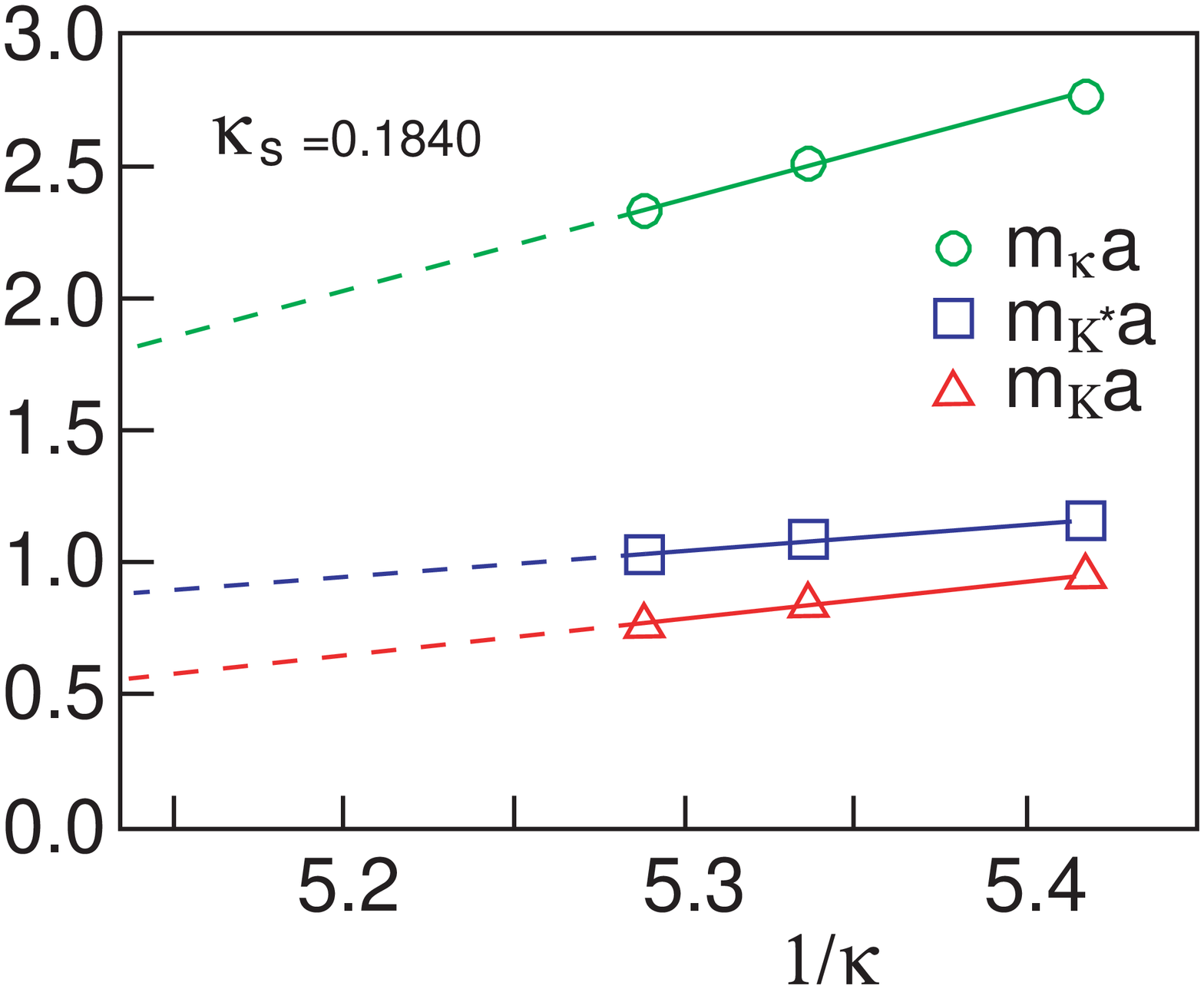} 
\end{minipage}
\caption{
Left :  $m_\rho$, $m_\sigma$ and $m_\pi^2$ in the lattice unit as functions of
the inverse hopping parameter. 
Right : 
$m_\kappa$, $m_{K^*}$ and $m_K$ in the lattice unit as functions of
the inverse hopping parameter. 
The $s$ quark hopping parameter is $\kappa_s=0.1840$.
}
\label{fig:Mass}
\end{center}
\end{figure}

From our results, we evaluate the critical value of the hopping parameter 
$\kappa_c = 0.195(3)$ and lattice space $a = 0.207(9)$ fm 
(CP-PACS has obtained $\kappa_c = 0.19286(14)$ and $a = 0.197(2)$ fm). 
Figure \ref{fig:Mass} (left)
shows masses of $\rho,\sigma$ and $\pi$ as  functions of $1/\kappa$.
We find that $m_\sigma/m_\rho$ at the chiral limit is  $0.33\pm 0.09$.

Finally, we present our preliminary result for $\kappa$ meson 
using the same common configurations. 
The $s$  quark is treated as a valence, {\it i.e.}, is used only 
in the propagator (\ref{Eq:propa-kappa}), not as a sea quark.
We adopt the same hopping parameter values, 
$\kappa=0.1846$, $0.1874$ and $0.1891$ for $u$ and $d$ quarks. 
We calculate three values of the hopping parameter for the $s$ quark:
$\kappa_s$ = 0.1835, 0.1840 and 0.1845. 
For each $\kappa_s$, we calculate masses of $\kappa$, $K^{*}$ and $K$
mesons, and extrapolate them to the chiral limit.  
Then we tune the $s$ quark hopping parameter, $\kappa_s$, 
to give the best experimental values for $m_{K^{*}}$ and $m_K$.
Figure \ref{fig:Mass} (right)
shows $m_\kappa a$, $m_{K^{*}} a$ and $m_K a$ as  functions of $1/\kappa$
for $\kappa_s=0.1840$.
Our preliminary analysis shows that value of $m_\kappa/m_{K^*}$ 
at the chiral limit is around $2.0$.

\vspace{-1.0cm}


\section{Concluding Remarks}

We have reported our exploratory study of the scalar mesons
based on the full QCD lattice  simulation with dynamical fermions.
Our results indicate the existence of a light $\sigma$
in the region $m_\pi < m_\sigma \le m_\rho$.   

An interesting observation is that 
the disconnected part gives a significant contribution, 
and this diagram makes the $\sigma$ meson light. 
This cannot be accessed in the framework of 
the non-relativistic quark model.
This point should be kept in mind in future phenomenological analyses
of the sigma meson, and also in lattice studies.
Note that the
$\kappa$ meson does not have such a mechanism, and therefore 
these two scalar meson masses can be different. 
The $\kappa$ and the valence $\sigma$ (connected part)
have the same structure in their propagators, but $s$ quark
mass is heavier than those of $u$ and $d$.  Therefore 
$m_\kappa > m_{conn}$, where $ m_{conn}$ is a mass corresponding
to the connected part, $\sigma_V$.

The calculations reported here have limitations discussed in Sect.1.
We expect that
these defects will be gradually overcome.  In lattice calculation,
once someone establishes the scale of simulation required to obtain 
meaningful results, many improved and large-scale works succeed.
Therefore in a few years, the lattice study of scalar mesons will
provide important and fundamental information to deepen our
understanding of hadron physics.

\noindent
{\bf Acknowledgment}
This work is supported by Grants-in-Aid for Scientific Research by
Monbu-Kagaku-sho (No.\ 11440080, No.\ 12554008, No.\ 12640263 and
No.\ 14540263),
DOE grants DE-FG02-96ER40495 
and ERI of Tokuyama Univ.
Simulations were performed on SR8000 at IMC, Hiroshima
Univ., SX5 at RCNP, Osaka Univ., and SR8000 at KEK.

\vspace{-4mm}

\end{document}